\documentclass[slac_one]{revtex4}
\usepackage{graphicx}
\usepackage{fancyhdr}
\def\Missing#1#2{{\mbox{$#1\kern-0.57em\raise0.19ex\hbox{/}_{#2}$}}}
\def\vMissing#1#2{\ifmmode
            \vec{#1}\kern-0.57em\raise.19ex\hbox{/}_{#2}
         \else
            {{\mbox{$\vec{#1}\kern-0.57em\raise.19ex\hbox{/}_{#2}$}}}
         \fi}
\def\lsim{\mathrel{\rlap{\lower4pt\hbox{\hskip1pt$\sim$}}
    \raise1pt\hbox{$<$}}}        
\def\gsim{\mathrel{\rlap{\lower4pt\hbox{\hskip1pt$\sim$}}
    \raise1pt\hbox{$>$}}}        
\def\met{\mbox{$\Missing{E}{T}$}}

\begin{document}

\title{{\small{Hadron Collider Physics Symposium (HCP2008),
Galena, Illinois, USA}}\\ 
\vspace{12pt}
QCD, Tevatron results and LHC prospects} 

%

\author{V.~Daniel Elvira for the D\O\ and CDF Collaborations}
\affiliation{FNAL, Batavia, IL 60510, USA}

\begin{abstract}
We present a summary of the most recent 
measurements relevant to Quantum Chromodynamics (QCD) delivered 
by the D\O\ and CDF Tevatron experiments by May 2008. CDF and D\O\ are moving 
toward precision measurements of QCD based on 
data samples in excess of 1~fb$^{-1}$. The inclusive jet cross sections
have been extended to forward rapidity regions and measured with unprecedented precision 
following improvements in the
jet energy calibration. Results on dijet mass distributions, 
$b\overline{b}$ dijet production using tracker based
triggers, 
underlying event in dijet and Drell-Yan samples, inclusive photon and diphoton cross 
sections complete the list of measurements included in this paper.
Good agreement with pQCD within errors is observed for jet production measurements. 
An improved and consistent theoretical description is needed for $\gamma$+jets processes.
Collisions at the LHC are scheduled for early fall 
2008, opening an era of discoveries at the new energy 
frontier, 5-7 times higher than that of the Tevatron.
\end{abstract}

\maketitle

\thispagestyle{fancy}


\section{PHYSICS MOTIVATION}

The last couple of years were filled with excitement at Fermilab. The 
Tevatron $p\overline{p}$ collider has operated at a center-of-mass 
energy of 1.96~TeV since 2001, a little less than 10$\%$ higher than in the 
1991-1995 first period of data taking. In Run~2, the Tevatron 
reached a peak luminosity of 2.85$\times$10$^{32}$~cm$^{-2}$sec$^{-1}$,
delivering more than 4~fb$^{-1}$, almost thirty times more data than collected 
in Run I. By the end of 2009, the expectation is to accumulate 
6-8~fb$^{-1}$ of data. This paper includes a summary of the most recent 
experimental measurements relevant to Quantum Chromodynamics (QCD) delivered 
by the D\O\ and CDF Collaborations by May 2008.

The measurement of the differential inclusive jet and dijet mass
cross sections in hadron collisions provides a direct test
of perturbative quantum chromodynamics
(pQCD). The high $p_T$ range is directly
sensitive to the strong coupling constant ($\alpha_s$) and the
parton distribution functions (PDFs) of the proton. Deviations
from pQCD predictions at large $p_T$ may indicate
new physical phenomena not described by the standard
model.

The underlying event (UE) is formed by the 
``beam-beam remnants'' from the breakup of the proton and antiproton. 
Experimentally, in a dijet sample, 
the UE is typically defined as everything in the event 
except the two outgoing hard scattered jets, and consists of the beam-beam 
remnants plus initial and final-state radiation. 
An accurate measurement of the UE is important to,
for example, understand particle (hadron) level measurements of jet cross 
sections
compared with parton and hadron level theoretical predictions.
The UE will be an important element of the hadronic environment 
at the LHC, affecting all processes from Higgs searches to physics beyond the 
standard model. It is therefore important to construct models to predict the 
UE at the LHC energies, based on the data currently available.

The production of a photon with associated jets in
the final state is a powerful probe of the dynamics of
hard QCD interactions. Different angular configurations
between the photon and the jets can be used
to extend inclusive photon production measurements and simultaneously test 
the underlying dynamics of QCD hard-scattering subprocesses in 
different regions of parton momentum fraction $x$ and large hard-scattering
scales $Q^2$.
Diphoton final states are a signature of many
interesting physics processes. The understanding of the QCD production 
mechanism is therefore a pre-requisite to a reliable search. For example, at 
the LHC, one of the main decay channels for the Higgs boson
would be the $\gamma\gamma$ final state. An excess production of 
$\gamma\gamma$
at high invariant mass could be a signature of large
extra dimensions. In many theories involving physics
beyond the standard model, cascade decays of heavy
new particles generate a $\gamma\gamma$ signature.

\section{SAMPLE SELECTION AND CORRECTIONS}

The D\O\ and CDF detectors are described elsewhere~\cite{d0det,cdfdet}.

\subsection{Jets}
The primary tool for jet detection in both D\O\ and CDF is the
calorimeter system which provides good electron and hadron energy resolution,
a fine segmentation, hermeticity, 
and shower containment. The tracker system plays a fundamental role in 
identifying the secondary vertex in the case of the b-jet cross section 
measurement.  

For the analyses shown in these proceedings, the Run II iterative seed-based 
cone jet algorithm including mid-points~\cite{jetalgo} (Midpoint algorithm) with cone 
radius 0.7 in rapidity $y$ and azimuthal angle is used to cluster energies 
deposited 
in calorimeter towers. The same algorithm is used for partons in the
pQCD calculations.

Cosmic rays and beam related backgrounds are removed by applying loose cuts
on the ratio of the event missing transverse energy (\met) and the leading jet
$E_T$ (D\O\ ), or on the \met$\,$ significance (CDF). Requirements on 
characteristics of shower development for genuine jets are used to
remove the remaining background due to electrons, photons,
and detector noise that mimic jets.

The jet $p_T$ is corrected for the energy response of the
calorimeter, energy showering in and out the jet cone,
and additional energy from event pile-up and multiple
proton interactions~\cite{d0inc,cdfinc}. The jet energy corrections fix the
calorimeter jet four-momentum to the particle (hadron) level energy.
The electromagnetic part of the calorimeter is calibrated
using $Z\rightarrow e^{+}e^{-}$. The $\eta$-dependence of the jet response 
is determined using event $p_T$ balance in dijet events. 
At D\O\ the $p_T$-dependent absolute 
correction is derived from event $p_T$ balance in $\gamma$-jet events. 
Further corrections due to the difference in response between
quark- and gluon-initiated jets are computed using
the {\sc pythia}\cite{pythia} event generator, passed through a 
geant-based~\cite{geant} simulation of the detector response. 
In CDF, the absolute correction is derived from Monte Carlo events, based 
on the {\sc geant3}~\cite{geant} detector simulation tool kit, in which a 
parameterized shower simulation, {\sc gflash}~\cite{gflash}, is used to 
model the energy deposited in the calorimeter. The {\sc gflash} parameters
are tuned to test-beam data for electrons, and high momentum charged pions and 
the in-situ collision data for electrons from Z decays and low-momentum charged
hadrons.
The fractional uncertainty
of the jet $p_T$ calibration is less than 2(3)$\%$ for D\O\ (CDF) in the 
kinematic range covered in the measurements.
  
Jet cross sections are unfolded to correct for the effect of finite 
energy resolution using a four-parameter ansatz function to parameterize the
$p_T$ dependence (D\O\ ), or a smeared {\sc pythia} distribution weighted to 
match the data (CDF)~\cite{d0inc,cdfinc}.

\subsubsection {b-jets}
Jets initiated by b-quarks are selected in CDF using a trigger 
based on two jets with $p_T>$20~GeV associated with 
two displaced tracks reconstructed using the Silicon Vertex 
Trigger (SVT) system. Offline, jets are tagged using an algorithm based
on the reconstruction of a secondary vertex inside the jet, and a requirement
for the impact parameter to be $>$120~$\mu$m. 
An SVT b-tagged jet is defined so that events with two such objects 
always pass the trigger. The efficiency for requiring two SVT b-tagged jets 
in an event is calculated from a Monte Carlo simulation.
The shape of the invariant mass of the tracks associated to the secondary 
vertex can be used to separate the contribution of heavy flavor jets from 
light quark and gluon jets. A two components fit to the data is performed, 
based on Monte Carlo templates. A "signal" template describes 
the $b\overline{b}$ case and a 
"background" template all the other possible contributions. 

\subsection{Photons}

The D\O\ experiment selects photon candidates
from clusters of calorimeter cells within a cone of radius
R=0.4 defined around a seed tower~\cite{d0det,d0phot1,d0phot2}. 
The final cluster energy is then re-calculated from
the inner cone with R=0.2.
The selected clusters are required to
have greater than 96$\%$ of their total energy contained
in the EM calorimeter layers. Isolated clusters are selected
by requiring the energy outside an R=0.2 cone to be a small 
fraction of the photon energy. 
The candidate EM cluster is required not to be spatially
matched to a reconstructed track. This is accomplished
by computing a $\chi^{2}$ function evaluating the consistency,
within uncertainties, between the reconstructed $\eta$ and
$\phi$ positions of the cluster and the closest track measured
in the layer located at the shower maximum position. The
corresponding $\chi^{2}$ probability is required to be $<$0.1$\%$.
Background contributions to the direct photon sample
from cosmic rays and from isolated electrons, originating
from the leptonic decays of $W$ bosons, important
at high $p^{\gamma}_{T}$, are suppressed with a cut on \met. 
Photons arising from decays of $\pi^{0}$ and $\eta$ mesons are
already largely suppressed by the requirements above,
and especially by photon isolation.
The position and width of the Z boson mass
peak, reconstructed from $Z\rightarrow e^{+}e^{-}$ events, are used
to determine the EM calorimeter calibration factors and
the EM energy resolution.
The CDF experiment uses a very similar selection criteria, described in
Ref.~\cite{cdfphot}.

\section{JET CROSS SECTIONS}

\subsection{Inclusive Jet Measurements}
\begin{figure}[htbp]
\centering
\includegraphics[width=160mm]{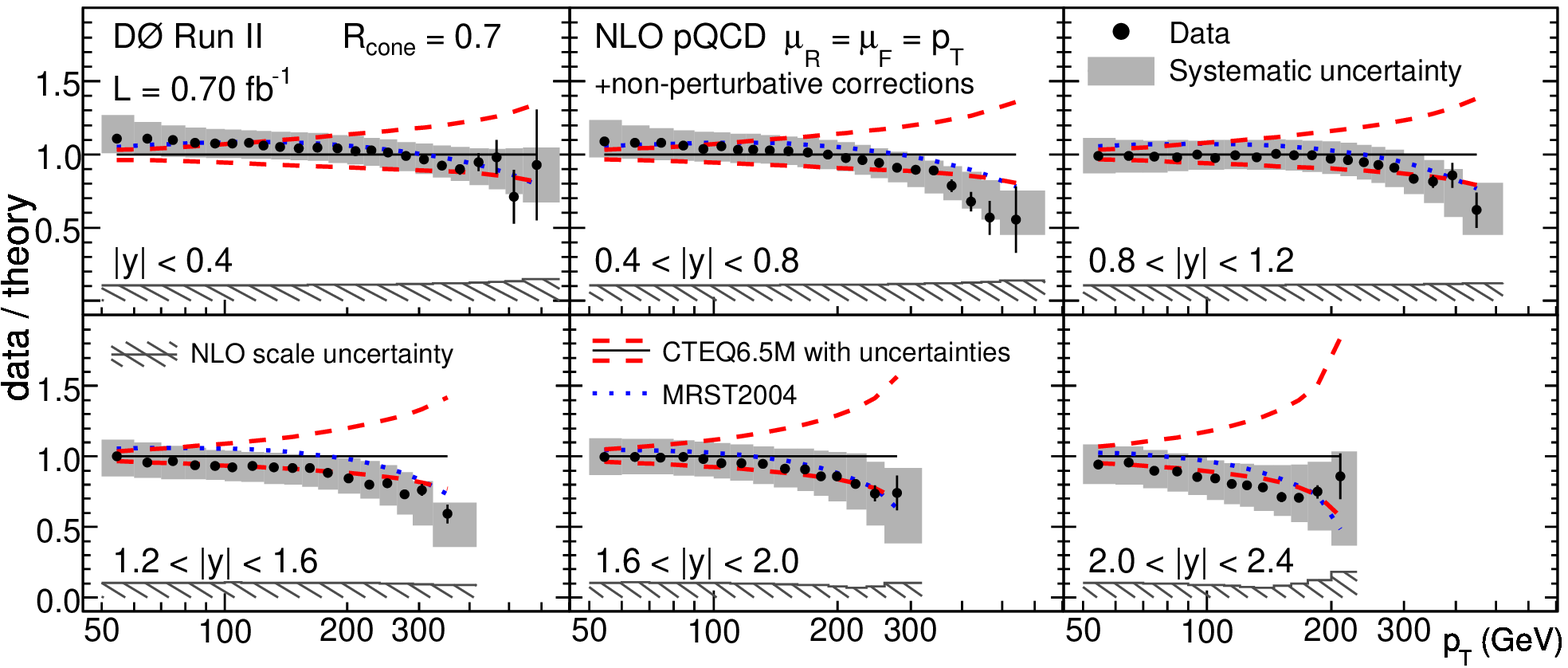}
\caption{Measured data-to-theory ratio for the inclusive jet cross section 
as a function of jet $p_T$ in six $|y|$ bins. 
The data systematic uncertainties are displayed by the full shaded band.} 
\label{fig:d0-incl-2}
\end{figure}

The D\O\ inclusive jet cross section measurements
corrected to particle (hadron) level are performed 
in six $|\eta|$ bins as a function of $p_T$ . The cross section corresponds
to an L=0.7~pb$^{-1}$ sample, extending
over more than eight orders of magnitude from $p_T$=
50~GeV to $p_T>$ 600~GeV. Perturbative QCD predictions
to next-to-leading order (NLO) in $\alpha_{s}$, computed using
the fastNLO program~\cite{fastnlo} (based on nlojet++~\cite{nlojet})
and the PDFs from CTEQ6.5M~\cite{cteq}, are compared to
the data. The predictions
are corrected for non-perturbative contributions
due to the underlying event and hadronization computed
by pythia with the CTEQ6.5MPDFs, the QWtune~\cite{tunes},
and the two-loop formula for $\alpha_{s}$. 
The ratio of the data to the theory is shown in Fig.~\ref{fig:d0-incl-2}.
The dashed lines show the uncertainties due to the different
PDFs coming from the CTEQ6.5 parameterizations.
The predictions from MRST2004~\cite{mrst} are displayed by
the dotted line. In all $y$ regions, the predictions agree
well with the data. There is a tendency for the data to
be lower than the central CTEQ prediction, particularly
at very large $p_T$, but the results are mostly within the CTEQ
PDF uncertainty band. The $p_T$ dependence of the data
is well reproduced by the MRST parameterization.
The point-to-point
correlations for the 24 different sources of systematic
uncertainties are given in Ref.~\cite{correl}.

\begin{figure}[htbp]
\centering
\includegraphics[width=120mm]{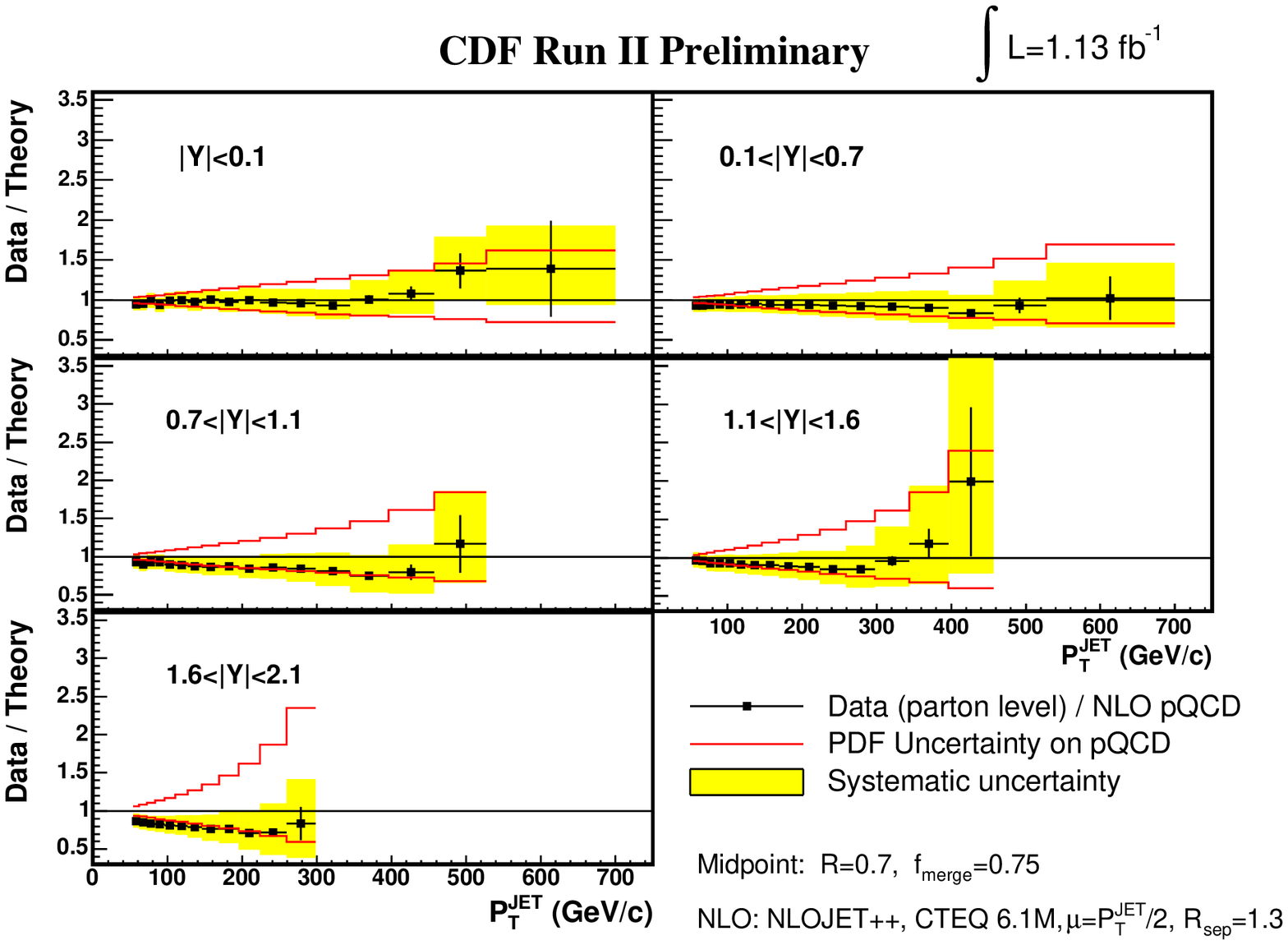}
\caption{Measured data-to-theory ratio for the inclusive jet cross section 
as a function of jet $p_T$ in six $|y|$ bins. 
The data systematic uncertainties are displayed by the full shaded band.} 
\label{fig:cdf-inc-2}
\end{figure}

CDF measured the inclusive differential jet cross sections as
a function of $p_T$ and rapidity, corrected to the hadron level.
The data-to-theory ratios, based on NLO pQCD predictions from
fastNLO~\cite{fastnlo}, are shown in
Fig.~\ref{fig:cdf-inc-2}. The measured inclusive jet cross sections tend
to be lower but still in agreement with the NLO pQCD
predictions.
To quantify the comparisons, a procedure based on a $\chi^{2}$ test
was performed which included information on individual systematic
uncertainties and their correlations, as well as different choices
in the theory calculation. This test yielded agreement 
probabilities of 71, 91, 23, 69, and 91$\%$ when
performed separately in the five rapidity regions.
CDF also measured the inclusive jet cross sections with a $k_T$ algorithm.
Reasonable agreement of the ratio of the $k_T$
and Midpoint cone results is reported in Ref.~\cite{cdfinc}.

\subsection{Dijet Mass Measurements}

\begin{figure}[htbp]
\centering
\includegraphics[width=120mm]{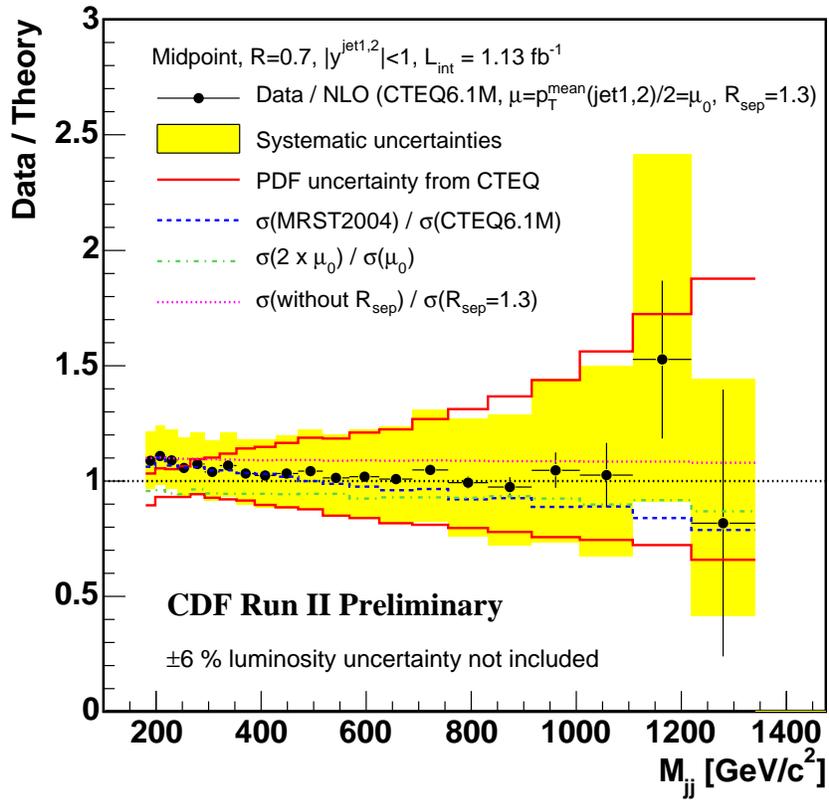}
\caption{The ratio of the CDF preliminary dijet mass cross 
section to NLO pQCD predictions, including the systematic error
band of the measurement, and the theory uncertainty from the choice of
PDFs.} 
\label{fig:cdf-dijets-2}
\end{figure}

A preliminary CDF dijet mass cross section measurement 
for jets with $|\eta|<$1 in a 
1.13~pb$^{-1}$ sample is compared in Fig.~\ref{fig:cdf-dijets-2} to the 
NLO pQCD predictions from 
fastNLO\cite{fastnlo,nlojet}. For the PDF in the proton, CTEQ6.1 is used,
and the renormalization and factorization scales are set to the average
$p_T$ of the leading two jets. The NLO pQCD predictions for jets of partons
are corrected for the non-perturbative underlying and hadronization effects. 
These are derived by running the jet clustering algorithm to the hadron on
parton and hadron level events generated with {\sc pythia}.
Good agreement between data and theory is observed to within the 
uncertainties in the measurement and the prediction.
An important motivation for this measurement is the search for new physics, 
which would show as deviations of the data with respect to the QCD 
predictions. Limits could be set, for example, to excited quark, 
massive gluon, $Z^{\prime}$ and $W^{\prime}$ production. We will not discuss
searches in these proceedings.

\subsection{$b\overline{b}$ Dijet Measurement}
\begin{figure}[htbp]
\centering
\includegraphics[width=120mm]{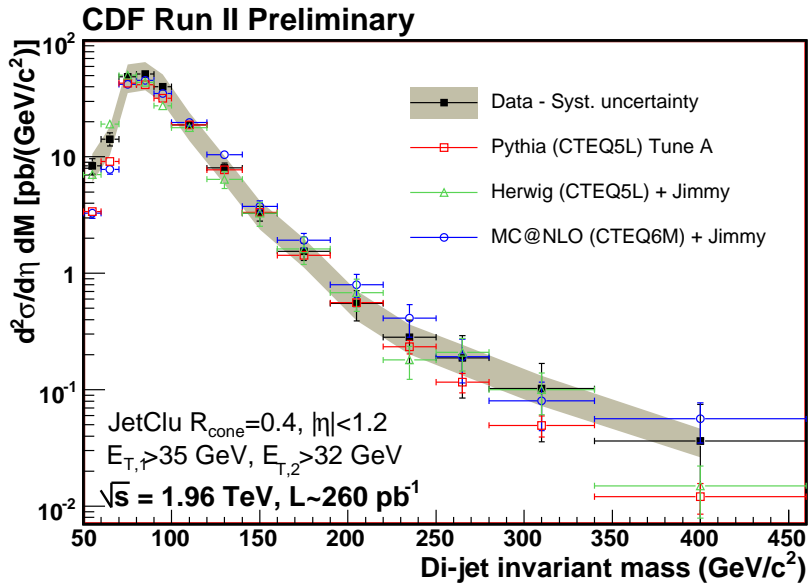}
\caption{Cross
section for $b\overline{b}$ dijet production as a function of the invariant
mass of the two jets.} 
\label{fig:cdf-bjet-mass}
\end{figure}

\begin{figure}[htbp]
\centering
\includegraphics[width=120mm]{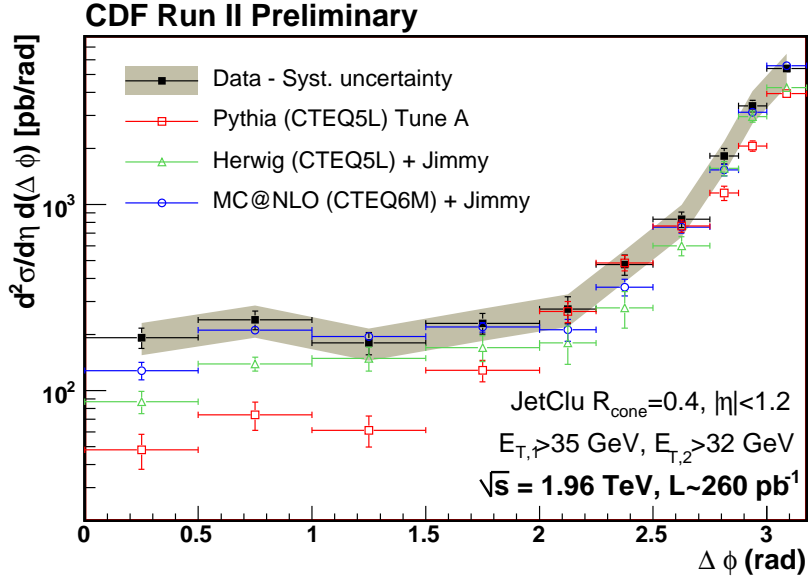}
\caption{Cross
section for $b\overline{b}$ dijet production as a function of the
separation in azimuthal angle between the two jets.} 
\label{fig:cdf-bjet-dphi}
\end{figure}

The CDF preliminary $b\overline{b}$ dijet cross section using the SVT is
based on a 260~pb$^{-1}$ sample.
Figures~\ref{fig:cdf-bjet-mass},~\ref{fig:cdf-bjet-dphi} show the cross
section for $b\overline{b}$ dijet production as a function of the invariant
mass and the separation in azimuthal angle of the two jets.
The measurement is compared to predictions from LO generators
such as {\sc pythia} and {\sc herwig}~\cite{herwig}, as well as to NLO
predictions from {\sc MC@NLO}\cite{mcnlo}, interfaced with the {\sc herwig}
parton shower and using a minimum quark 
$p_T$ of 10~GeV in $|\eta|<$1.75 and CTEQ6.1M. 
{\sc pythia} samples are generated 
with the ``tune A''~\cite{tunea} for the underlying event modeling. 
{\sc Jimmy}~\cite{jimmy} is used with {\sc herwig} and {\sc MC@NLO} to 
include the effect of multiple parton interactions. As illustrated in 
Figs.~\ref{fig:cdf-bjet-mass}-\ref{fig:cdf-bjet-dphi},
the agreement between the data and the theory improves as we move from 
a LO prediction to {\sc Herwig} or {\sc MC@NLO} with {\sc Jimmy}, 
which includes multiple parton interactions.

\section{UNDERLYING EVENT}

CDF has released a preliminary measurement of the
underlying event based on 2.7~fb$^{-1}$ samples. 
Although the UE is formally defined as the contributions from the
remnants of the colliding beams and multiple parton interactions, 
it is difficult to separate 
these contributions from those of initial and final state radiation. 
The UE is therefore studied from dijet events by measuring all particles 
except those associated with the two outgoing jets, and from Drell-Yan (DY) 
events by excluding the two outgoing leptons.
In dijet events, ``TransMax'' and ``TransMin'' regions, describes the two
sectors in $\phi$ perpendicular to the direction of the leading jet. This
definition allows to separate the hard from the soft component of the UE; 
while TransMax is sensitive to both the initial/final state radiation 
and the beam remnants, TransMin is more sensitive to the beam remnants.
In DY events, the ``Toward'' region is defined in the direction of the
di-lepton system originated from the $Z$ boson; the
``Transverse'' and ``Toward'' regions are therefore both sensitive to the 
UE, while the ``Away'' region contains the hadronic recoil.

\begin{figure}[htbp]
\centering
\includegraphics[width=120mm]{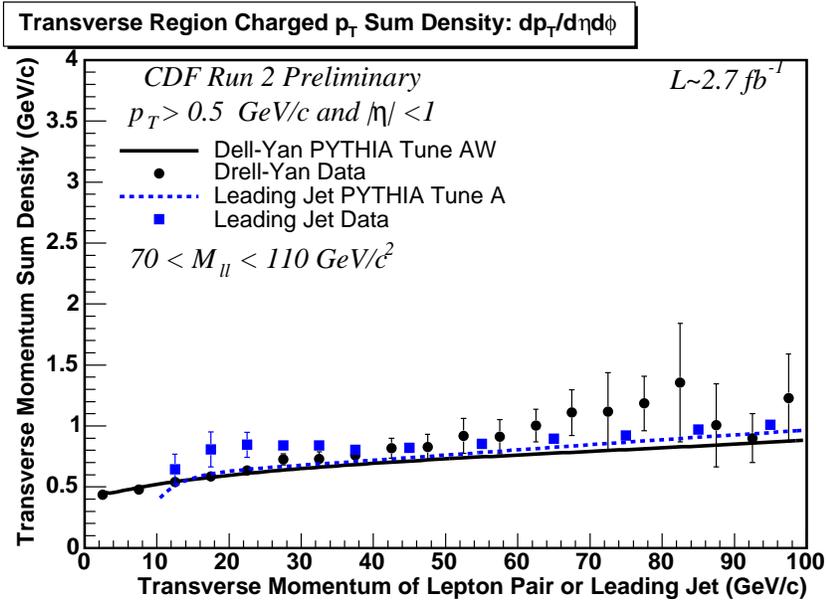}
\caption{Transverse momentum sum density as a function of the $p_T$ of
either the lepton pair or the leading jet. The results from the dijet and
the combined electron/muon sample are compared.
The error bars include both the statistical and systematic uncertainties.} 
\label{fig:cdf-ue-2}
\end{figure}

CDF measured the scalar $p_T$ sum density of the charged 
particles in DY events as a 
function of the transverse momentum of the lepton pair (electron and muon
samples combined). While the sum
increases with $p_T$ in the Away region due to the contribution of the
hadronic recoil, the Toward and Transverse regions show a flat behavior. 
The comparison of the Transverse regions for DY and dijet data in
Fig.~\ref{fig:cdf-ue-2} shows a 
similar trend, suggesting universality of the UE in 
hard scattering processes. The dotted lines are the different {\sc pythia} 
tunes which clearly describe the data reasonably well.

\section{PHOTON PRODUCTION}

\subsection{Inclusive $\gamma$+jets Measurement}
D\O\ has recently performed measurements of the inclusive $\gamma$ and
$\gamma$+jet cross sections~\cite{d0phot1,d0phot2}. The CDF Collaboration has
released a preliminary measurement of the inclusive $\gamma$ cross section based on
451~pb$^{-1}$.
The $\gamma$+jet+X final state is
dominated by compton $qg$ scattering for $p_T \lsim $120~GeV. 
The D\O\ measurement is based on a 1~fb$^{-1}$ sample and probes the gluon density function
in the  0.007$\le X \le$0.8 range for
900$\le Q^{2} \equiv$ $(p^{\gamma}_{T})^{2} \le$ 1.6$\times$ 10$^{5}$~GeV$^{2}$. 
The trick is achieved by measuring the 
cross section for different angular configurations between the $\gamma$ and the 
leading jet: central or forward, positive or negative $y^{\gamma}$ and $y^{jet}$ 
as a function of $p^{\gamma}_{T}$.

\begin{figure}[htbp]
\centering
\includegraphics[width=100mm]{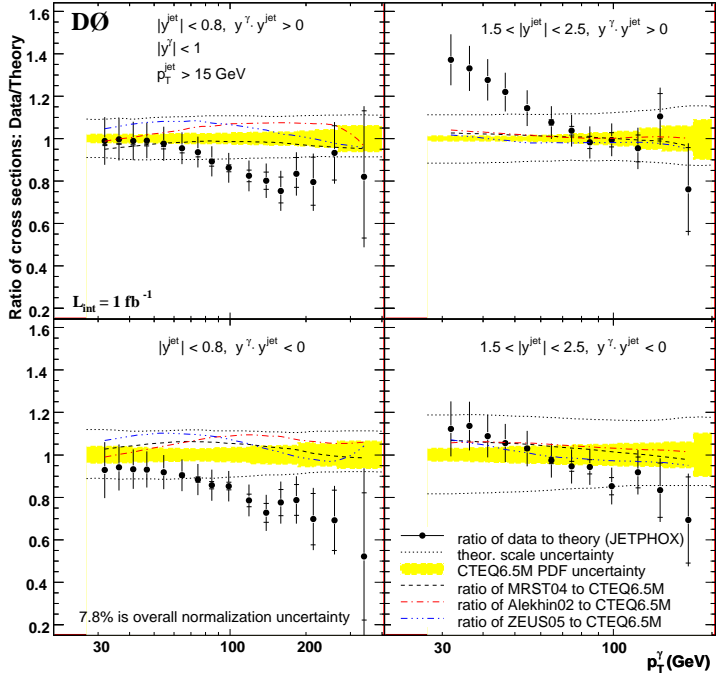}
\caption{The ratios of the measured cross section, in each measured interval, 
to the NLO QCD prediction using {\sc jetphox} with the CTEQ6.5M PDF set.} 
\label{fig:d0-phot-2-08}
\end{figure}

The ratio of the measured cross section to the NLO
QCD prediction {\sc jetphox}~\cite{jetphox} is taken in each interval and the results
are shown in Fig.~\ref{fig:d0-phot-2-08}. The
data-to-theory ratios have a shape similar to those observed
in the inclusive photon cross sections measured by the
UA2, CDF and D\O\ collaborations.
Different choices of PDFs or parameters in the theory
are not able to simultaneously accommodate the measured
differential cross sections in all of the regions. 

\begin{figure}[htbp]
\centering
\includegraphics[width=100mm]{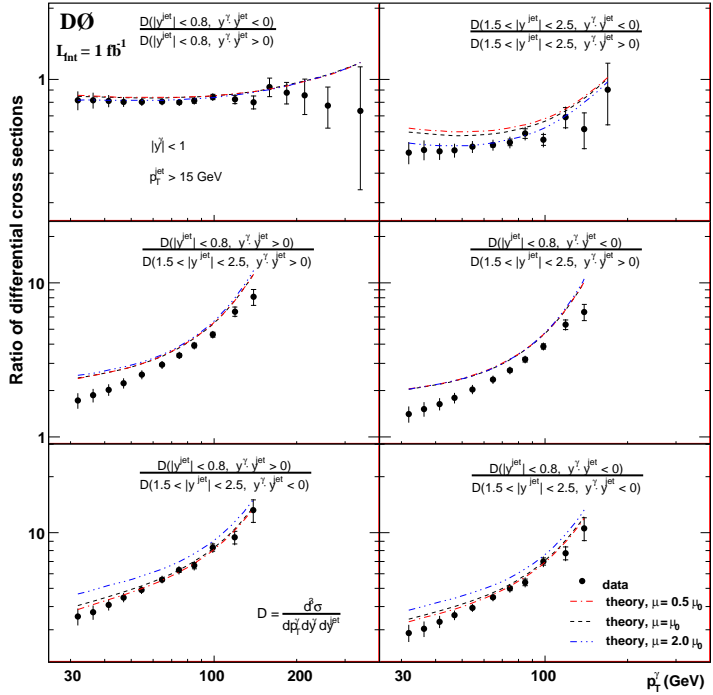}
\caption{The ratios between the cross sections in each $|y^{jet}|$ region. 
The solid vertical error bars correspond to the
statistical and systematic uncertainties added in quadrature while the horizontal 
marks indicate the statistical uncertainty.
NLO QCD theoretical predictions for the ratios are estimated using {\sc jetphox}.} 
\label{fig:d0-phot-3-08}
\end{figure}

A more precise measurement can be performed by taking the ratios of cross sections for 
different configurations, for which most of the systematic uncertainties cancel, 
leaving a residual 3-9$\%$ error across most of the $p_T$ domain. 
These ratios, shown in Fig.~\ref{fig:d0-phot-3-08}, are qualitatively reproduced by the 
theory.
A quantitative difference, however, is observed for the ratios of the central
jet regions to the forward 1.5$<|y^{jet}|<$2.5, $|y^{jet}|>$ 0
region, even after the theoretical scale variation is taken
into account.

\subsection{Diphoton Measurement}
\begin{figure}[htbp]
\centering
\includegraphics[width=120mm]{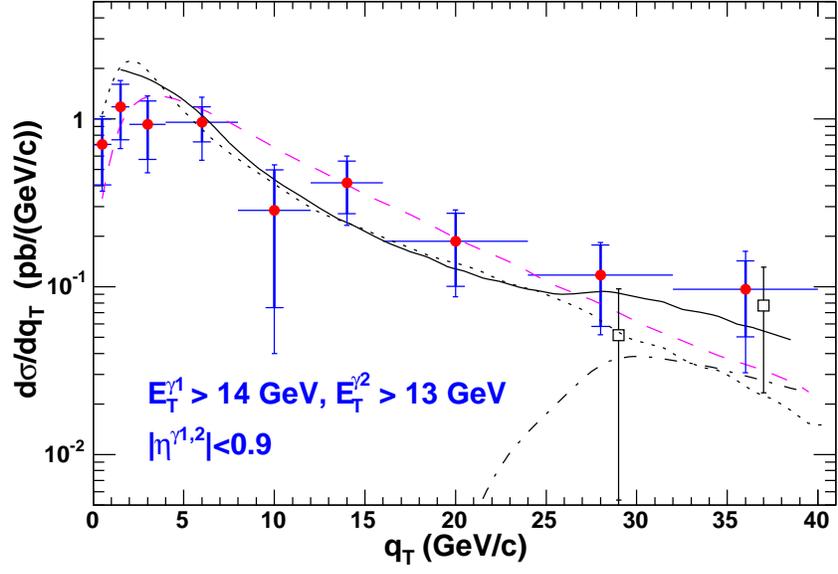}
\caption{The $\gamma\gamma$ $q_T$ distribution from the CDF
Run II data, along with predictions from {\sc diphox} (solid line),
{\sc resbos} (dashed line), and {\sc pythia} (dot-dashed line).} 
\label{fig:cdf-diphot-2}
\end{figure}

\begin{figure}[htbp]
\centering
\includegraphics[width=120mm]{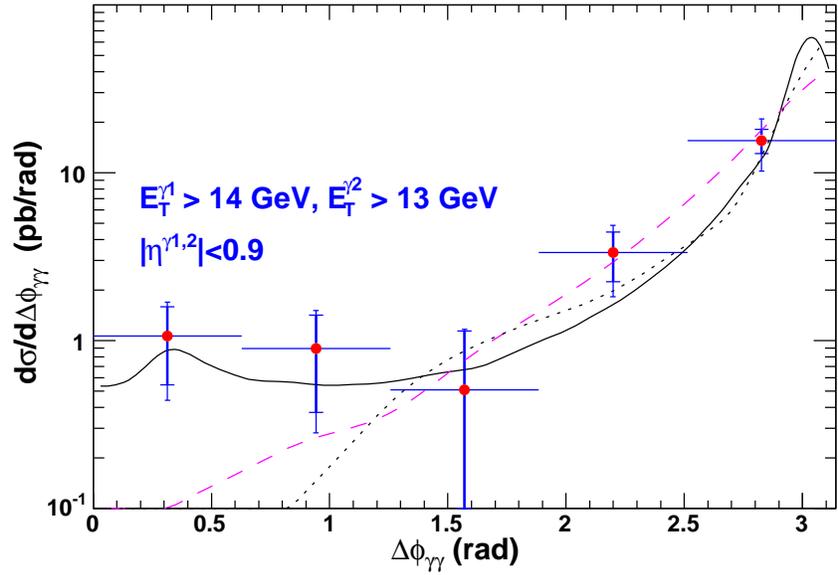}
\caption{The $\gamma\gamma$ $\Delta \phi_{\gamma\gamma}$
angular distribution from the CDF
Run II data, along with predictions from {\sc diphox} (solid line),
{\sc resbos} (dashed line), and {\sc pythia} (dot-dashed line).} 
\label{fig:cdf-diphot-3}
\end{figure}

A result which is unique to CDF is the measurement of diphoton distributions. 
The two dominant contributions to diphoton production
come from the LO $q\overline{q}$ process which is dominant at high mass and the 
NLO $gg$ process which, in spite of being suppressed by
$\alpha_{s}^{2}$ with respect to the $q\overline{q}$ diagram, 
is still relevant at low mass. 
The measurement is based on a 207~pb$^{-1}$ sample, with
889 diphoton candidate events surviving the 
selection requirements. 
Results are compared to the
LO {\sc pythia} calculation, 
the {\sc diphox}~\cite{diphox} NLO prediction, which includes the $gg$ process, and
{\sc resbos}~\cite{resbos}, a NLO calculation which resums the effects of 
initial state soft 
gluon radiation but includes only LO fragmentation contributions. 
Figures~\ref{fig:cdf-diphot-2},~\ref{fig:cdf-diphot-3} show the diphoton cross sections 
as a 
function of the photon $q_T$ and $\Delta \phi_{\gamma\gamma}$. 
It is apparent that the data favors the {\sc resbos} calculation at low $q_T$ and 
$\Delta \phi_{\gamma\gamma}$ greater than $\phi/2$, where initial state gluon 
radiation is important.
By contrast, in the region where fragmentation becomes relevant, 
large $q_T$ and $\Delta \phi_{\gamma\gamma}<\pi$/2, {\sc diphox} does a better job.
For agreement in all areas, a resummed full NLO calculation would be  necessary.

\section{SUMMARY AND LHC PROSPECTS}

The Tevatron experiments are entering an era of precision QCD measurements based on 
samples in excess of 1~fb$^{-1}$.
Good agreement with pQCD within errors is observed for jet production measurements. 
An improved and consistent theoretical
description is needed for $\gamma$+jets. Collisions at the LHC are scheduled for early fall 
2008, opening an era of discoveries at the new energy 
frontier, 5-7 times higher than that of the Tevatron.





\end{document}